\documentclass[12pt]{JHEP3}
\usepackage{epsfig}
\epsfclipon
\usepackage{multicol}
\usepackage{epsfig,bm}
\usepackage{amssymb,amsmath}
\usepackage{graphicx}

\newcommand{\be}{\begin{equation}}
\newcommand{\ee}{\end{equation}}
\newcommand{\bea}{\begin{eqnarray}}
\newcommand{\beas}{\begin{eqnarray*}}
\newcommand{\eea}{\end{eqnarray}}
\newcommand{\eeas}{\end{eqnarray*}}
\newcommand{\ba}{\begin{array}}
\newcommand{\ea}{\end{array}}
\def\ls{\mathrel{\lower4pt\vbox{\lineskip=0pt\baselineskip=0pt
           \hbox{$<$}\hbox{$\sim$}}}}
\def\gs{\mathrel{\lower4pt\vbox{\lineskip=0pt\baselineskip=0pt
           \hbox{$>$}\hbox{$\sim$}}}}

\newcommand{\e}{\textrm{e}}

\newcommand{\vp}{\varphi}

\newcommand{\varm}{\varphi_{\max}}
\newcommand{\fnl}{f_{NL}^{\phi}}

\def\smiley{\hbox{\large$\bigcirc$\hspace{-.80em}%
\raise.2ex\hbox{$\cdot\cdot$}\kern-.61em    
\lower.2ex\hbox{\scriptsize$\smile$}}\ }

\newcommand{\roughly}[1]{\mathrel{\raise.3ex\hbox{$#1$\kern-0.85em
\lower1ex\hbox{$\sim$}}}}

\newcommand{\lsim}{\roughly<}

\def\be{\begin{equation}}
\def\beq\begin{equation}
\def\ee{\end{equation}}
\def\bea{\begin{eqnarray}}
\def\eea{\end{eqnarray}}

\def\beq{\begin{equation}}
\def\eeq{\end{equation}}
\def\beqa{\begin{eqnarray}}
\def\eeqa{\end{eqnarray}}

\newcommand{\bmat}{\left(\begin{array}}
\newcommand{\emat}{\end{array}\right)}

\title{Cosmological constraints on string scale and coupling
arising from tachyonic instability}

\author{Kari Enqvist$^{1,2}$, Asko Jokinen$^{3}$, Anupam Mazumdar$^{3}$,
Tuomas Multam\"aki$^{3}$, Antti V\"aihk\"onen$^{1,2}$,\\
 ${}^1$ Helsinki Institute of Physics, P.O. Box 64,
FIN-00014 University of Helsinki, Finland\\
 ${}^2$ Department of Physical Sciences,
P.O. Box 64, FIN-00014 University of Helsinki, Finland\\
 ${}^3$ NORDITA, Blegdamsvej-17, Copenhagen, DK-2100, Denmark}

\abstract{We demonstrate that string motivated inflation ending via
tachyonic instability leaves a detectable imprint on the cosmic
microwave background (CMB) radiation by virtue of the excitation of
non-Gaussian gravitational fluctuations. The present WMAP bound on
non-Gaussianity is shown to constrain the string scale by $M_S/M_P\leq
10^{-4}$ for string coupling $g_{s}<0.1$, hence improving the existing
bounds. If tachyon fluctuations during inflation are not negligible,
we find the stringent constraint $g_s\sim 10^{-9}$ for
$M_{S}/M_{P}<10^{-3}$. This case may soon be ruled out by the
forthcoming CMB non-Gaussinianity bounds.}

\preprint{NORDITA-2005-18\\ HIP-2005-07/TH}
\keywords{Non-Gaussianity, Tachyons, Strings, Inflation }

\begin{document}


\section{Introduction and Summary}

Currently the main hope for detecting signatures of string theory
relies on the cosmic laboratory. In particular, the cosmic microwave
background (CMB) signals of an early inflationary phase, which could
have a stringy origin. One possibility discussed in the literature is
a period of accelerated expansion induced by the dynamics of two
parallel branes~\cite{DvaliTye,Cliff1,Others,Maz}. A D-brane $Dp$ and
an anti-D-brane $\widetilde{Dp}$ approach each other, giving rise to
inflation, which ends when the branes annihilate, or equivalently by
virtue of a tachyonic instability that arises when the branes approach
each other to within the string length~\cite{Cliff1,Others,Maz}. Here
the tachyon is a collective description of excited open string degrees
of freedom representing the breaking of space time supersymmetry and
the resulting vacuum instability. Tachyons are also found in bosonic
string theory and in the context of a non-BPS brane
configurations~\cite{Joebook}, suggesting that their appearance is
quite common in string theory.

The process of $Dp$-$\widetilde{Dp}$ annihilation and the associated
tachyon dynamics can be depicted in terms of an effective scalar field
theory with a negative mass squared potential~\cite{Sen}. However, it
has turned out to be extremely difficult to get enough e-foldings of
inflation in string motivated inflationary models in general.  Both
well studied $Dp$-$\widetilde{Dp}$ inflation~\cite{Maldacena} and the
race track models~\cite{Cliff3} have a difficulty in generating a
large enough number of e-foldings, about $N\sim 65$. If inflation
occurs in a number of subsequent bouts, it is however possible to
relax~\cite{Cliff2} the bound on $N$, and the minimum number of
e-foldings could be as low as $12$ in order to explain the large scale
structure and the CMB data; another $20$ to $30$ e-foldings are needed
to solve the flatness problem. However, here we do not commit
ourselves to one particular view but rather simply assume that enough
e-foldings will be obtained and that eventually inflation ends via a
tachyonic instability. As we will show, then there exists a direct
test of string cosmology: the tachyonic instability gives rise to
non-Gaussianity in the primordial perturbations (for a review of
non-Gaussianity, see \cite{ngrev}).  Moreover, such non-Gaussian
features can easily be observable and, by virtue of the WMAP
data~\cite{WMAP-nongaussian}, are already constrained.

In the string theory setup, the tachyon field rolls down from the top
of the potential after the end of inflation. As the field rolls down
quickly through the region where the effective mass squared is still
negative, the first order matter perturbations grow exponentially
similarly to the case of well-studied tachyonic
preheating~\cite{Linde}~\footnote{The tachyon(s) could roll slowly as
well, which would lead to tachyon(s) driven inflation, for instance
see Ref.~\cite{Maz}, in which case the non-Gaussianity parameter can
be estimated with the help of slow roll parameters~\cite{cal}, which
is typically small.}. Following our previous analysis,~\cite{we2}, one
then finds that the growth in the matter perturbations seed the second
order metric fluctuations\footnote{ This is a generic feature of
preheating, see~\cite{we1}.} and gives rise to large non-Gaussianity
in the CMB temperature fluctuation spectrum. As will be discussed
below, the present observable limits~\cite{WMAP-nongaussian} on
non-Gaussianity can then be used to constraint the string scale,
$M_{S}$, and the string coupling, $g_{s}$.

We consider two particular scenarios. In both the cases the initial
fluctuations of the tachyon matter is created during
inflation. However in order to keep our analysis as general as
possible, we consider the first case where the first order metric
perturbations, $\phi^{(1)}={\rm constant}$, has a trivial solution,
which is dominated by the amplitude of the existing perturbations
during inflation.  In this case we obtain a bound on the string scale:
$M_S/M_P\leq 10^{-4}$ for $g_s\leq 0.1$. In the second scenario we
assume the growing solution for $\phi^{(1)}$, in which case we obtain
a stringent constraint on the string coupling, given by $g_{s}\sim
10^{-9}$ for $M_S/M_P\sim 10^{-3}$. Future observational limits on
non-Gaussianity can soon render the latter scenario incompatible with
the observed CMB temperature fluctuation amplitude of $10^{-5}$.


\section{Gravitational perturbations and the tachyon}

For simplicity we focus on the bosonic action for the tachyon. The
discussion can be easily generalized to the superstring case. The 4D
effective field theory action of the tachyon field $T$ on a D3-brane,
computed in a bosonic string theory around the top of the potential,
reads as (up to higher derivative terms)
\cite{Gerasimov:2000zp,Kutasov:2000aq}
\begin{equation}
\label{tachyonaction} S_B = -\tau_3 \int d^4x \sqrt{-g}
\left(\alpha' e^{-T}
\partial_{\mu} T \partial^{\mu} T
 + (1+T) e^{-T} \right) \,,
\end{equation}
where the brane tension is given by
\begin{equation}
\label{tension} \tau_3 = \frac{M_S^4 \sqrt{2}}{(2\pi)^3 g_s}.
\end{equation}
Here $g_s$ is the string coupling and $M_S=1/l_s = 1/\sqrt{\alpha'}$
are the fundamental string mass and length scales.

The form of the tachyon action given in Eq.~(\ref{tachyonaction}) is
not immediately suitable for the purposes of perturbation equations at
second order. It is convenient first to transform the action into a
canonical form by the redefinition

\begin{equation}
\label{reparam2} T = -2 \, \textrm{ln} \left(
\frac{\varphi}{2\sqrt{2\tau_3\alpha'}} \right).
\end{equation}
The potential for the redefined tachyon $\varphi$ reads as
\begin{equation}
\label{repot} V(\varphi) = \frac{1}{8} \, M_S^2 \varphi^2 \left[ 1
- 2 \, \textrm{ln} \left( \frac{\varphi}{\varphi_{\max}}
\right)\right],
\end{equation}
where the definitions of $\tau_3$ and $\alpha'$ have been used and
\begin{equation}
\label{maximum} \varphi_{\max} \equiv 2 \sqrt{2\tau_3\alpha'} =
M_S \sqrt{\frac{\sqrt{2}}{\pi^3\,g_s}} \,.
\end{equation}
The transition $T=0\rightarrow\infty$ corresponds to $\varphi$
evolving from the maximum of the potential at $\varphi_{\max}$ to
$\varphi=0$.  The tachyonic instability is present when
$V''(\varphi) < 0$ or $1/\e < \varphi/\varphi_{\max} \leq 1$.
Because of the instability, the occupation number of $\varphi$
quanta grows exponentially.

In the simplest brane anti-brane inflationary scenarios the inflaton
is the modulus $\sigma$, which is governed by the brane anti-brane
separation or the angular separation between the two branes. In
either case the inflaton potential can be separated from the tachyon
potential, such that the total potential is given
by~\cite{Cliff1,Others},
\begin{equation}
V=V(\sigma)+V(\varphi)\,.
\end{equation}
The potential of $V(\sigma)$ depends on twice the brane tensions
$2\tau_3$ and $\sigma$. In this paper we do not have to know the exact
form of $V(\sigma)$, which can be found in Refs.~\cite{Cliff1,Others}.
However for the purpose of simplifying our analysis we demand that
after the end of inflation, when the tachyon starts rolling, the VEV
of the inflaton is vanishing, $\langle \sigma\rangle =0$.  This can be
achieved by proper redefinition of the modulus around the string scale
$M_{S}$.

The tachyon field and the inflaton field can be divided into the
background and the first and the second order perturbation after
inflation:
\begin{eqnarray}
\label{split} \varphi &=& \varphi_0(\eta) +
\delta^{(1)}\varphi(\eta,{\bf x}) + \frac{1}{2}
\delta^{(2)}\varphi(\eta,{\bf x})\,. \\
\sigma &=& \delta^{(1)}\sigma(\eta,{\bf x}) + \frac{1}{2}
\delta^{(2)}\sigma(\eta,{\bf x}) \,.
\end{eqnarray}
Here $\eta$ denotes the conformal time. The field perturbation gives
rise to the metric perturbation (for details see Ref.~\cite{ngrev} and
references therein), which up to the second order reads $\phi =
\phi^{(1)} + \phi^{(2)}/2$ with \cite{Acquaviva}
\begin{eqnarray}
\label{metric}
g_{00} &=& - a(\eta)^2 \left( 1 + 2\phi^{(1)} + \phi^{(2)} \right)\,,\\
g_{0i} &=& 0 \,,\\
g_{ij} &=& a(\eta)^2 \left( 1 - 2\psi^{(1)} - \psi^{(2)} \right)
\delta_{ij}\,,
\end{eqnarray}
where the generalized longitudinal gauge is used and the vector
and tensor perturbations are neglected. Here $a(\eta)$ is the scale
factor.
The background equations of motion are then found to be
\begin{eqnarray}
\label{eqom}
3 {\cal H}^2 &= &\frac{1}{2M_{P}^2} \varphi_0'^{\,2} +
\frac{a^2}{M_{P}^2} V(\varphi_0) \,,\\
0&=& \varphi_0'' + 2 {\cal H} \varphi_0' + a^2 V'(\varphi_0)\,,
\end{eqnarray}
while the $\sigma_b$-equation is trivial. Here $\cal H$ denotes the
Hubble expansion rate expressed in conformal time and the reduced
Planck scale is given by $M_P=2.4\times 10^{18}$~GeV. The relevant
first order perturbation equations can be written in the
form~\cite{Antti,we1,we2}
\begin{eqnarray}
\phi^{(1)\,''} - \partial_i \partial^i \phi^{(1)} + 2 \left( {\cal H} -
  \frac{\varphi_0''}{\varphi_0'} \right) \phi^{(1)\,'}  + 2
\left( {\cal H}' - \frac{\varphi_0''}{\varphi_0'} {\cal H} \right) \phi^{(1)}
&=& 0 \,.\label{bardeeneq}\\
\delta^{(1)}\sigma'' + 2 {\cal H} \delta^{(1)}\sigma' - \partial_i \partial^i
\delta^{(1)}\sigma + g^2 \varphi_0^2 \, \delta^{(1)}\sigma &=& 0\,.
\label{chieq}
\end{eqnarray}
All the information regarding $\delta^{(1)}\varphi$ is contained in
Eq.~(\ref{bardeeneq}), whose right hand side is zero by virtue of
$\langle \sigma\rangle =0$, see \cite{MFB}. Further note that there
are no metric perturbations in Eq.~(\ref{chieq}). This is due to
assuming a vanishing VEV for $\sigma$. Hence the $\sigma$ part can be
solved separately.

At the second order we are only interested in the gravitational
perturbation arising due to the tachyon and the inflaton field.  The
gravitational perturbation equation can be written in an expanding
background as~\cite{Antti,we1,we2}
\begin{eqnarray}
{\phi^{(2)''}} + 2 \left({\cal H}-\frac{\vp''_0}{\vp'_0}\right)
{\phi^{(2)'}} + 2 \left({\cal H}'- \frac{\vp''_0}{\vp'_0}{\cal
H}\right) {\phi^{(2)}} - \partial_i \partial^i{\phi^{(2)}} = \nonumber \\
{\cal J}_{\varphi, \textrm{local}} + {\cal J}_{\sigma, \textrm{local}}+
{\cal J}_{\textrm{non-local}} \label{eq:phi2}
\end{eqnarray}
where the source terms $\cal J$ are quadratic combinations of first order
perturbations; in particular \cite{Antti,Acquaviva},
\begin{eqnarray}
\label{local0}
  {\cal J}_{\varphi, \textrm{local}} &=& -\frac{1}{M_P^2}
\left[ 2(\delta^{(1)}\varphi')^2
+ 8(\varphi_0')^2 (\phi^{(1)})^2 - a^2 V_{,\varphi\varphi}
(\delta^{(1)} \varphi)^2 - 8 \varphi_0' \phi^{(1)}
\delta^{(1)}\varphi' \right]\nonumber \\
&&- 24{\cal H}' (\phi^{(1)})^2 - 24 {\cal H} \phi^{(1)}\phi^{(1)\,'}\\
{\cal J}_{\sigma, \textrm{local}} &=& -\frac{2}{M_{P}^2} (\delta^{(1)}
\sigma')^2 +  \frac{a^2}{M_{P}^2} \frac{\partial^2 V}{\partial \sigma^2}
(\delta^{(1)} \sigma)^2~\,, \\ 
{\cal J}_{\textrm{non-local}} &=& \triangle^{-1}
f(\delta^{(1)}\varphi,\delta^{(1)}\sigma,\phi^{(1)})\,,
\end{eqnarray}
where $f$ is a quadratic function of the first order fluctuations and
the coefficients depend on background quantities. Because of the
inverse Laplacian the last source term is non-local.  Typically such
term contains: $\triangle^{-1}(\phi^{(1)\,'}\triangle\phi^{(1)})$,
$\triangle^{-1}(\partial_i\delta^{(1)}\varphi \partial^i
\delta^{(1)}\varphi),\ldots$ Note that the left hand side of
Eq.~(\ref{eq:phi2}) is identical to the first order equation, see
Eq.~(\ref{bardeeneq}). $\cal J_{\textrm{non-local}}$ involves an
inverse spatial Laplacian, thus rendering it non-local~\footnote{The
measure of non-Gaussianity is the non-linearity parameter
$\phi^{(2)}=f_{NL}^{\phi} (\phi^{(1)})^2$. In general $f_{NL}^{\phi}$
contains momentum dependent part, i.e. $f_{NL}^{\phi}({\bf k_1},{\bf
k_2})$, and the constant piece. It is the non-local terms which affect
the momentum dependent part, since all the derivatives are replaced by
momenta in the Fourier space. However the present constraint on
non-Gaussianity parameter from WMAP does not give the momentum
dependent constraint but only the constant part. Therefore the
non-local terms do not lead to any observable constraints, so we do
not consider them here.}.


\section{Estimating non-Gaussianity}

Let us assume that the unstable modes grow within a time interval
much smaller than the Hubble rate in Eq.~(\ref{bardeeneq}). This
means that the tachyon  field evolves fast enough so that we can
neglect the effects of the expansion. Then conformal time and
cosmic time are equal with $\eta=t$ and $a=1$ and in the large
wavelength limit we then obtain
\begin{equation}
\label{bardeeneq2}
\ddot\phi^{(1)} - 2A\, \dot\phi^{(1)} = 0\,,
\end{equation}
where $A=\ddot\varphi_0/\dot\varphi_0$. With the assumption that the
tachyonic stage is so short so that $\dot\varphi_0,\,\ddot\varphi_0$
are effectively constants, there will be two solutions: a constant
$\phi^{(1)}$; and an exponentially growing solution $\phi^{(1)}
\propto \exp(2A t)$.  The amplitudes of the two solutions are
determined by the initial conditions.

In the brane-anti-brane case, the dynamics of $\varphi$ is decoupled
from inflation; therefore no instability occurs during the initial
phase of inflation.  During which the tachyon fluctuations is also
suppressed and negligible. However as the branes move close to each
other the tachyonic instability is triggered. Let us assume that the
scale of inflation is such that $H_{inf}\sim M_{S}$ during the last
${\cal O}(5-10)$ e-foldings of inflation, then the tachyon feels the
vacuum induced long wavelength fluctuations during inflation.  Note
that these fluctuations are isocurvature in nature. Let us first
investigate the scenario where $\phi^{(1)}={\rm constant}$. This
assumes that no amplification in $\phi^{(1)}$ has taken place due to
the rolling tachyon. By virtue of the observed temperature anisotropies
we take $\phi^{(1)}\sim 10^{-5}$, as we shall see that this conclusion
does not hold for higher order metric perturbations. They will be
amplified in order to give rise to a significant non-Gaussianity.

When the brane separation becomes of the order of string length, the
mass squared of $\varphi$ field becomes negative. We assume this
happens when the inflaton $\sigma \rightarrow 0$.  During this epoch
the fluctuations of the tachyon field grow exponentially for all the
momenta $k<M_{S}/\sqrt{2}$. The dispersion of the growing modes at $t>0$ is
given by~\cite{Linde}
\begin{equation}
\langle \delta \varphi^2\rangle
=\int_{0}^{M_S/\sqrt{2}}\frac{dk^2}{8\pi^2}
e^{2t\sqrt{M_S^2/2-k^2}}=
\frac{e^{\sqrt{2}M_St}(\sqrt{2}M_St-1)+1}{16\pi^2t^2}\,. 
\label{dispersion}
\end{equation}
The growth of the fluctuations continues until the fluctuations can
modify the background equations of motion. This happens when
Eq.~(\ref{dispersion}) reaches $\langle\delta\varphi\rangle
\sim\sqrt{\varphi_{max}/e}$, or until the time span $t_{\ast}\sim
(\sqrt{2}/M_S)\ln(4\pi\varphi_{max}/M_S)$. During this period the
number density of the tachyon quanta is then given by~\cite{Linde}
\begin{equation}
n_{k}\sim e^{\sqrt{2}M_St_{\ast}}\sim \frac{16\sqrt{2}}{\pi g_s}\,.
\end{equation}
In the limit when $g_s$ becomes vanishingly small the number density
of excited tachyonic quanta diverges. This phenomena is easy to
understand. In the small coupling limit the brane tension becomes
large. Therefore a large potential energy density becomes available to
the excited quanta. A similar situation is found in the tachyonic
preheating case~\cite{Linde, we2}.

The total energy density stored in the produced $\varphi$ quanta
is given by
\begin{equation}
\label{energy2}
\rho_{\varphi}\sim \frac{1}{2} (\delta^{(1)}\dot\varphi)^2
 \sim \frac{M_S^4}{3B^3 g_s}\,.
\end{equation}
where $B=\pi / \sqrt{2}$. The equation for the second order metric
perturbation now includes a source term which includes perturbations
from the inflaton, ${\cal J}_{\sigma, \textrm{local}}$ and ${\cal
J}_{\varphi, \textrm{local}}$. However when the tachyon is rolling the
inflaton fluctuations are negligible compared to that of the tachyon.
This allows us to keep only the tachyonic source terms, which in the
long wavelength regime reads as
\begin{equation}
\label{phi22} \ddot\phi^{(2)} \sim -
\frac{4M_S^4}{3B^3 g_s M_P^2}\,.
\end{equation}
Integrating the above equation over the time interval
$t_{\ast}\sim(\sqrt{2}/M_S)\ln(4\pi\varphi_{max}/M_S)$, we find
$\phi^{(2)}\sim - (4/3B^3g_s)(M_S/M_P)^2\ln^2(4/\sqrt{Bg_s})$.
An important point to note here is that the second order metric
perturbations, $\phi^{(2)}$, are mainly sourced by the tachyon matter
fluctuations.

The non-Gaussianity parameter during the rolling of the tachyon, when
the first order metric perturbation stays constant, is then roughly
given by
\begin{equation}
\label{fnltach}
f_{NL}^{\phi}\sim \frac{\phi^{(2)}}{(\phi^{(1)})^2} \sim
-\frac{4\cdot 10^{10}}{3B^3g_s}
\left(\frac{M_{S}}{M_{P}}\right)^2\ln^2\left
(\frac{4}{\sqrt{B g_s}}\right)\,.
\end{equation}
The parameter $f_{NL}^{\phi}$ is related to the standard
non-Gaussianity parameter $f_{NL}$ by $f_{NL}=-f_{NL}^{\phi}+11/6$
\cite{ngrev}. The present bound from WMAP is given by the range
$-132<f_{NL}^{\phi}<60$ $@~95\%$~\cite{WMAP-nongaussian}, In Fig. 1 we
have plotted the $f_{NL}^{\phi}=-132$ curve based on
Eq.~(\ref{fnltach}). The allowed parameter space is below the curve.


%
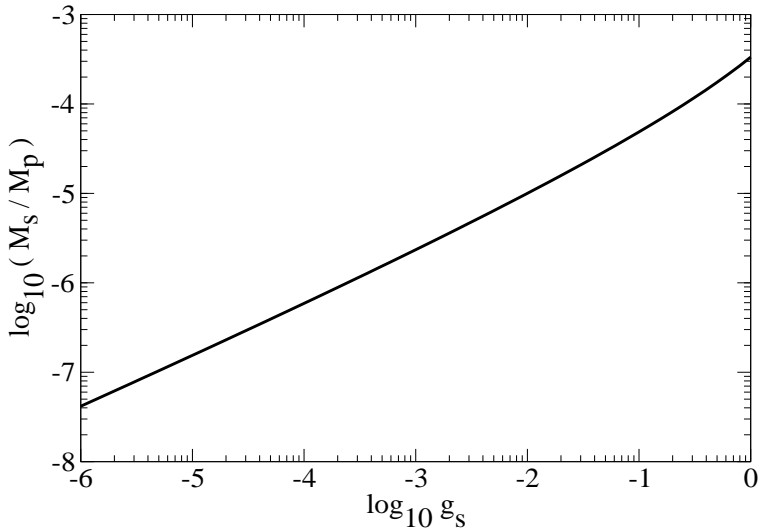
\EPSFIGURE{stringtachyon1.eps,width=10cm,height=7cm} {Allowed region is
below the curve, which gives $f_{NL}^{\phi}=-132$.}
%


It has been argued that an unavoidable consequence of brane inflation
is a network of cosmic strings that appears at the end of inflation
\cite{Sarangi}.  If true, and assuming that the amount of cosmic
strings produced is the highest possible allowed by the current CMB
data, which is less than $10\%$ of the total energy density, then the
CMB and LSS data would imply a bound $G\mu<10^{-6}$ on the string
tension $\mu$~\cite{Pogosian} (here $G$ is the Newton's constant). A
search of cosmic strings in the WMAP 1-year data has been shown to
lead to the limit $G\mu<10^{-6.5}$~\cite{Smoot}. For a fixed dilaton,
these bounds translate to bounds on the string scale, given
respectively by $M_{S}/M_{P}<10^{-3}$ and $M_{S}/M_{P}<10^{-3.25}$.

Limits from non-Gaussianity, as given by Fig.~1, yield an improved
constraint on the string scale if we require that the string coupling
$g_{s}< 1$. For instance, if $g_{s}\sim 10^{-1}$, as is appropriate
for the grand unified theories, we find a bound on the string scale
given by $M_{S}/M_{P}\leq 10^{-4}$.


\section{Consequences of tachyon fluctuations during inflation}

Let us now consider the other extreme case, when the first order
metric perturbation would be dominated by the exponentially growing
solution so that $\phi^{(1)} \propto \exp(2A t)$. This can happen when
the initial fluctuations in $\varphi$ is large enough to overcome the
constant solution for $\phi^{(1)}$.  In this case the bound on the
string scale and the string coupling will be more stringent than the
one obtained in the previous Section, as we will now discuss.

Following Eq.~(\ref{eq:phi2}) we may write
\begin{eqnarray}
\label{bardeeneq3}
 \ddot\phi^{(2)} - 2A\,\dot\phi^{(2)} = -\frac{1}{M_P^2}
\left[ 2(\delta^{(1)}\dot\varphi)^2
+ 8\dot\varphi_0^2 (\phi^{(1)})^2 - V_{,\varphi\varphi}
(\delta^{(1)} \varphi)^2 - 8 \dot\varphi_0 \phi^{(1)}
\delta^{(1)}\dot\varphi \right],
\end{eqnarray}
where we assumed that the right hand side is mainly sourced
by the exponential instability triggered by the tachyon.
Note that $\delta^{(1)}\varphi$ can be solved through the Einstein
constraint~\cite{ngrev}
\begin{equation}
\label{eincon}
\delta^{(1)}\varphi = \frac{2M_P^2}{\dot\varphi_0}
\left(\dot\phi^{(1)} + H \phi^{(1)} \right)\,,
\end{equation}
which is valid only if the backround value of the second scalar field
vanishes $\sigma_b=0$. As discussed in~\cite{we2}, now $\phi^{(2)}$
contains a homogeneous solution $\sim \exp(2At)$ together with a
source part $\sim \exp(4At)$. After a while the source part dominates
and we obtain for the non-Gaussianity parameter,
\begin{equation}
\label{param}
f_{NL}^{\phi}= \frac{\phi^{(2)}}{(\phi^{(1)})^2} = 8 +
\frac{M_P^2 2V_{,\varphi\varphi}}{\dot\varphi_0^2}
- \frac{4 M_P^2 \ddot\varphi_0^2}{\dot\varphi_0^4}
- \frac{\dot\varphi_0^4}{M_P^2\ddot\varphi_0^2} \,.
\end{equation}
Here we neglect a possible but small time variation in $A$.

The region of tachyonic growth occurs near $\varphi_0\approx \varm$
so that we obtain
\begin{equation}
\label{param22}
f_{NL}^{\phi} \approx 8 - 16 C g_s \left( \frac{M_P}{M_S} \right)^2 -
\frac{1}{C g_s} \left( \frac{M_S}{M_P} \right)^2 - 2C g_s \left(
\frac{M_P}{M_S} \right)^2 \rm{ln}^2 \left( \frac{4\pi}{(C~g_s)^{1/2}}
\right)\,,
\end{equation}
where $C= \pi^3/\sqrt{2}$.  It is easy to see that $\fnl$ is bounded
from above by $\fnl\lsim 8$, which is comfortably within the current
experimental bounds. On the other hand, there is no lower bound on
$\fnl$, allowing us to set constraints on the values of $M_S/M_P$
and $g_s$. In Fig. 2 we show the allowed region (between the lines)
for different values of $\fnl$. The current WMAP lower bound
$-132<\fnl$ $@~95\%$ c.l.~\cite{WMAP-nongaussian} corresponds to the
contour given by the solid line in Fig. 2.

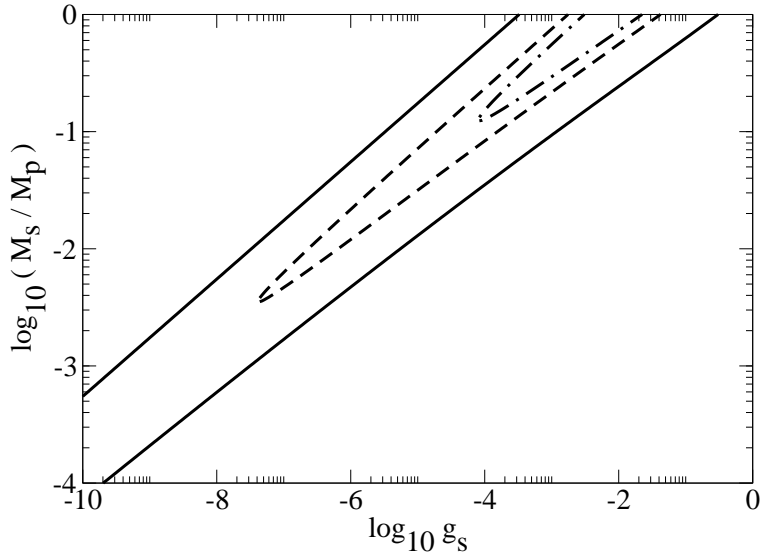
\EPSFIGURE{stringtachyon2.eps,width=10cm} {The contours given by the
solid, dashed and dot-dashed lines correspond the non-linearity
parameters $f_{NL}^{\phi}=-132,\, -20, \, -10$ respectively.}

The non-Gaussianity constraint on the string coupling $g_s$ can be
seen to be quite stringent.  For $M_S/M_P\leq 10^{-3}$, from Fig.~2
one finds $g_s\sim 10^{-9}$. A future, improved bound on
$f_{NL}^{\phi}$ would move the ratio $M_S/M_P$ towards one, rendering
such models incompatible with the CMB observations because a too large
anisotropy would be imparted onto the temperature
fluctuations. Further note that the present considerations can be used
to rule out low scale brane-anti-brane inflation, which would give
rise to a large non-Gaussianity parameter.

In summary, if there is a large isocurvature component during
inflation due to the fluctuations of the tachyon, then
brane-anti-brane inflation ending via tachyon condensation is in great
troubles. The present level of observational sensitivity to
non-Gaussianity can already rule out a large part of the parameter
space and requires a very small string coupling.  In such a case the
challenge for string cosmology would be to build brane-anti-brane
inflationary models with $H\ll M_{S}$. For such a large $M_S$, the
isocurvature perturbations of the tachyon will be very much suppressed
during inflation, thereby evading our bounds depicted in Fig.~2.
Otherwise when the isocurvature component is small then we can improve
the obtained bound on the string scale $M_{S}/M_{P}\leq 10^{-4}$.

Note that our conclusions hold for a tachyonic instability occurring
in bosonic strings, but a similar analysis can be performed for the
superstring case. Moreover, although we considered a particular
scenario of brane-anti-brane inflation, our arguments can be applied
also to a tachyonic instability occurring in non-BPS branes. Thus
non-Gaussianity may prove to be a vital tool for constraining string
motivated toy models of inflation.


\vskip40pt

We would like to thank Cliff Burgess, Andrew Liddle, Horace Stoica,
David Lyth and Filippo Vernizzi for discussions. A.V.~is supported by
the Magnus Ehrnrooth Foundation. A.V.~thanks NORDITA and NBI for their
kind hospitality during the course of this work. K.E. is supported in
part by the Academy of Finland grant no. 75065.

\vskip 30pt

\end{document}